\documentclass[a4paper,fleqn,usenatbib]{mnras}

\usepackage{ae,aecompl}

\usepackage{graphicx}	
\usepackage{amsmath}	
\usepackage{amssymb}

\title[]{Probing the {\color{red}H}e~II re-{\color{red}I}onization {\color{red}ER}a via {\color{red}A}bsorbing {\color{red}C}~IV {\color{red}H}istorical {\color{red}Y}ield ({\color{red}HIERACHY}) I: A Strong Outflow from a $z\sim4.7$ Quasar}

\author[Yu et al.]{Xiaodi Yu$^{1}$,
Jiang-Tao Li$^{2}$\thanks{E-mail: pandataotao@gmail.com},
Zhijie Qu$^{2}$, 
Ian U. Roederer$^{2,3}$,
Joel N. Bregman$^{2}$,
 \newauthor Xiaohui Fan$^{4}$,
 Taotao Fang$^{1}$, 
 Sean D. Johnson$^{2}$, 
 Feige Wang$^{4}$, 
 and Jinyi Yang$^{4}$\\
$^{1}$Department of Astronomy, Xiamen University, Xiamen, Fujian 361005, China.\\
$^{2}$Department of Astronomy, University of Michigan, 311 West Hall, 1085 S. University Ave, Ann Arbor, MI, 48109-1107, U.S.A.\\
$^{3}$Joint Institute for Nuclear Astrophysics -- Center for the Evolution of the Elements (JINA-CEE), USA.\\
$^{4}$Steward Observatory, University of Arizona, 933 North Cherry Avenue, Tucson, AZ 85721, USA.\\
}

\date{Accepted XXX. Received YYY; in original form ZZZ}

\pubyear{2021}

\begin{document}
\label{firstpage}
\pagerange{\pageref{firstpage}--\pageref{lastpage}}
\maketitle

\begin{abstract}
Outflows from super-massive black holes (SMBHs) play an important role in the co-evolution of themselves, their host galaxies, and the larger scale environments. Such outflows are often characterized by emission and absorption lines in various bands and in a wide velocity range blueshifted from the systematic redshift of the host quasar.
In this paper, we report a strong broad line region (BLR) outflow from the $z\approx4.7$ quasar BR~1202-0725 based on the high-resolution optical spectrum taken with the Magellan Inamori Kyocera Echelle (MIKE) spectrograph installed on the 6.5m Magellan/Clay telescope, obtained from the ``Probing the He II re-Ionization ERa via Absorbing C IV Historical Yield'' (HIERACHY) project. This rest-frame ultraviolet (UV) spectrum is characterized by a few significantly blueshifted broad emission lines from high ions;
the most significant one is the \ion{C}{IV} line at a velocity of $\sim-6500 \rm~km~s^{-1}$ relative to the H$\alpha$ emission line, which is among the highest velocity BLR outflows in observed quasars at $z>4$. The measured properties of UV emission lines from different ions, except for \ion{O}{I} and Ly$\alpha$, also follow a clear trend that higher ions tend to be broader and outflow at higher average velocities.
There are multiple \ion{C}{IV} and \ion{Si}{IV} absorbing components identified on the blue wings of the corresponding emission lines, which may be produced by either the outflow or the intervening absorbers.
\end{abstract}


\section{Introduction} \label{sec_intro}

In the past decade, super-massive black holes (SMBHs) with masses $>10^{9}~M_{\odot}$, which are comparable to the most massive ones in the local Universe, were detected in the epoch of reionization (z>6) \citep[e.g.,][]{Wu15,Inayoshi19,Yang20,Wang21}.
How these monsters grow so fast in the early Universe, and why they stopped growing after an initial fast growth stage (e.g., \citealt{Trakhtenbrot11}) are fundamental problems which could provide unique insights on the co-evolution of SMBHs, host galaxies, and their environments.


The growth of a SMBH is tightly linked to the growth of its host galaxy, as suggested by the strong correlation between the nuclear black hole (BH) mass and the velocity dispersion of galactic bulge \citep[e.g.,][]{Gebhardt00,KL13}.
Although the gravity of a SMBH is negligible compared to its host galaxy, it could modulate the physical state of the gas reservoir surrounding the host galaxy by injecting energy and momentum via various kinds of feedback \citep[e.g.,][]{Fabian12,King15,He19}. 
These feedback processes have been extensively studied by detecting nuclear outflows through extremely blueshifted ultraviolet (UV) emission and absorption lines of luminous quasars \citep[e.g., ][]{Trump06,Rich11,King15,He17}.
Broad high-ionization UV emission lines (e.g., \ion{C}{IV} $\lambda\lambda1548,1550$ \AA), that show blueshift relative to the low-ionization ones, are detected in many quasars \citep[e.g.,][]{Gaskell82,Van01,Rich02,Wang11,Sun18}. These broad emission lines are emitted from gas within the broad line region (BLR). Therefore, such blueshift indicates that outflows may be a common configuration in the BLR \citep[e.g.,][]{Gaskell82,Corb96,Mar96,Zam02,Rich11,Den12}. Despite the vast number of observations, the acceleration mechanisms for these BLR outflows and their connections to outflows detected in absorption are still poorly understood. High spectral resolution and signal-to-noise ratio ($S/N$) rest-frame UV spectra of quasars are useful to study these questions, providing high sensitivity in detecting and dynamically resolving the weak absorption features from outflowing clouds. These features have been used to study the structure and dynamics of outflows \citep[e.g.,][]{Ham11,CC18}.

Our group is composing a high spectral resolution sample of the brightest quasars at $z>4.5$. It is primarily designed for the project HIERACHY, which probes the He II re-ionization era at $z\gtrsim3$ via \ion{C}{IV} absorption lines, but is also informative to study BLR outflows.
As the first paper of this project, we herein report a high resolution optical spectrum of a $z\approx4.7$ quasar BR~1202-0725 obtained with the 6.5m Magellan telescope.
BR 1202-0725 was discovered as a quasar candidate in the APM\footnote{the Automatic Plate Measuring facility in Cambridge.}
multicolor survey \citep[e.g.,][]{Irwin91} and  
further confirmed as a quasar at $z\sim4.69$ with the strong Ly$\alpha$ emission line \citep{Storrie96}.
Recently, a more reliable measurement of the redshift of $z = 4.689\pm0.005$ was obtained with the H$\alpha$ emission line \citep{Jun15}.
As one of the UV brightest quasars at $z>4$ \citep[$m_{\rm 1450(1+z_{em}),AB}= 18.00$,][]{Storrie96}, BR~1202-0725 hosts a massive SMBH \citep[$M_{\rm SMBH}\sim 10^{10}~M_{\odot}$,][]{Jun15} and shows a significant velocity difference between the \ion{C}{IV} and Ly$\alpha$ emission lines \citep[][]{Storrie96}. 
Besides, BR~1202-0725 also resides in a system undergoing the gas-rich-major-merger, and is located in one of the most overdense fields known in the early universe. This system consists of a quasar, a submillimeter galaxy, and (at least) three Ly$\alpha$ emitters within $\lesssim 10''$ \citep[corresponding to $\lesssim 70$ kpc,][]{Drake20}.
BR~1202-0725, its host galaxy and galaxy environment have been observed in a broad wavelength range to study their multi-phase atomic gas traced by Ly$\alpha$, [\ion{O}{I}] 145.5$\mu \rm m$, [\ion{C}{I}] 370$\mu \rm m$, [\ion{O}{II}] $\lambda3727~$\AA, [\ion{C}{II}] 158$\mu \rm m$ and [\ion{N}{II}] 122,205$\mu \rm m$ \citep[e.g.,][]{Ohta00,Salome12,Wagg12,Car13,Lag18,Lee19,Drake20,Lee21}, molecular gas traced by CO, HCN and $\rm HCO^{+}$ \citep[e.g.,][]{Salome12,Jones16,Lee21}, as well as dust and star formation activities traced by FIR emission \citep[e.g.,][]{Omont96,Iono06,Salome12}.

The present paper is organized as follows: in \S\ref{sec_obs}, we introduce the observations and data reduction. In \S\ \ref{emlpro}, we estimate the properties of the SMBH and measure various emission or absorption line components likely associated with a BLR outflow. We further compare the properties of the BLR outflow of BR 1202-0725 to those of other quasar-driven outflows from literature. Our main results are summarized in \S\ \ref{sec_summary}. Throughout the paper, we assume a $\Lambda$CDM cosmology with a Hubble constant of $H_0=70 \rm~km~s^{-1}$ Mpc$^{-1}$ and cosmological parameters of $\Omega_m=0.3$, 
and $\Omega_\Lambda=0.7$. All the errors are quoted at 1$\sigma$ confidence level.

\section{Observations and Data Reduction} \label{sec_obs}
We observed BR 1202-0725 on 2019 April 27 
with the high-resolution echelle spectrograph MIKE \citep[][]{Ber03}
installed on the 6.5m Landon Clay (Magellan II) Telescope (with detailed information summarized in Table~\ref{obs}). 
MIKE is a double echelle spectrograph, achieving a spectral resolving power of $R \sim 32,000$ in the red channel (4800-9400~\AA) and $R \sim 41,000$ in the blue channel (3350-5000~\AA) 
with a $0\farcs7\times5\farcs0$ entrance slit and a $2 \times 2$ binning of the detector.
With the exposure time of $\sim$ 3~hours ($6\times1800$ seconds), the $S/N$ at 1320, 1400 and 1510 \AA~(rest-frame) are around 35, 36 and 40, respectively,
which enables us to identify and measure \ion{C}{IV} absorption features with column density as low as log$(N_{\rm C IV}/\rm cm^{-2}) \sim 12.3$.

The raw data is reduced with the MIKE dedicated pipeline CarPy \citep{Ke00,Ke03} 
\footnote{ https://code.obs.carnegiescience.edu/mike}, 
which includes the overscan subtraction, pixel-to-pixel flat field division, image coaddition, cosmic ray removal, sky and scattered-light subtraction, rectification of the tilted slit profiles along the orders, spectrum extraction, and wavelength calibration.
We use the default CarPy settings for extracting the spectrum and
IRAF \citep[][]{Tody93} to perform the flux calibration with the standard star LTT 3218 observed in the same night.

\begin{figure*}
\centering
\includegraphics[width=1.\textwidth]{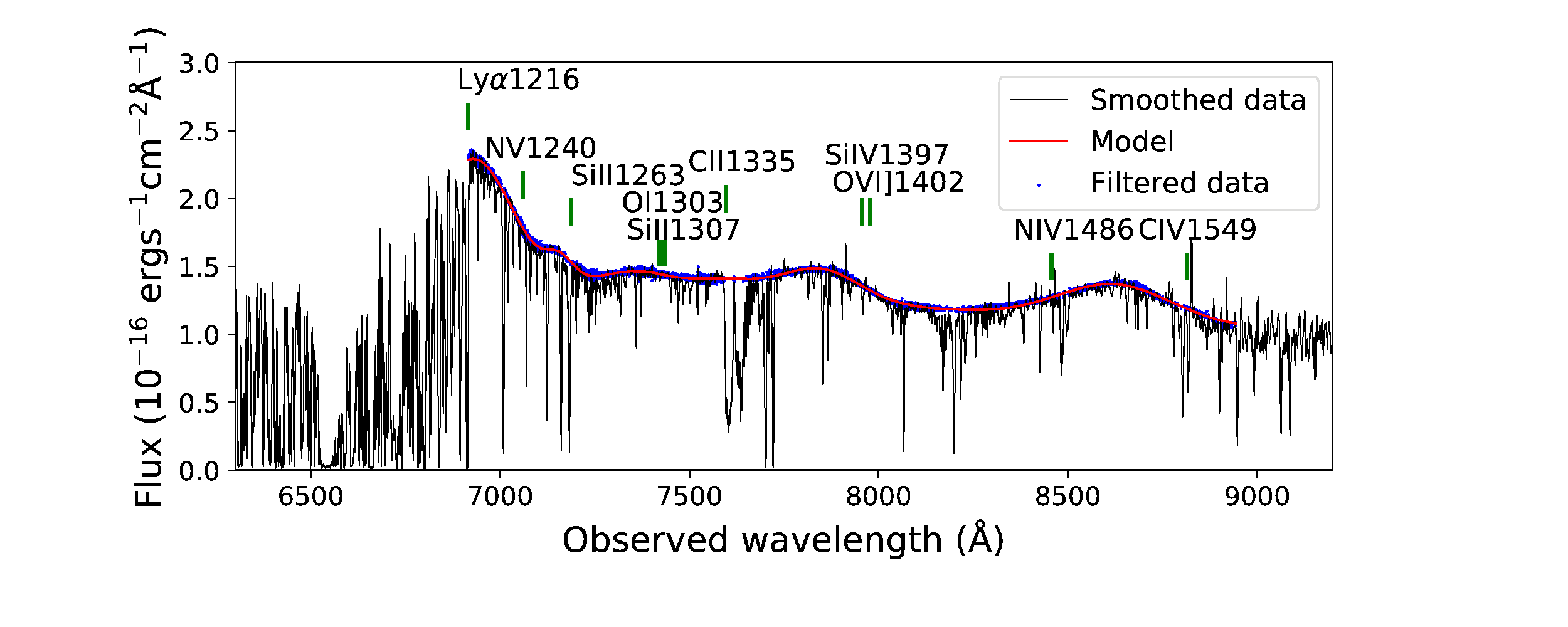}
\caption{Smoothed MIKE spectrum of BR 1202-0725 (black). The filtered featureless spectrum and the best-fit model are plotted as blue dotted and red solid curves, respectively.
The expected positions of emission lines at the quasar systematic redshift are shown in green vertical bars.
\label{fs}}
\end{figure*}

\begin{figure*}
\centering
\includegraphics[width=1.\textwidth]{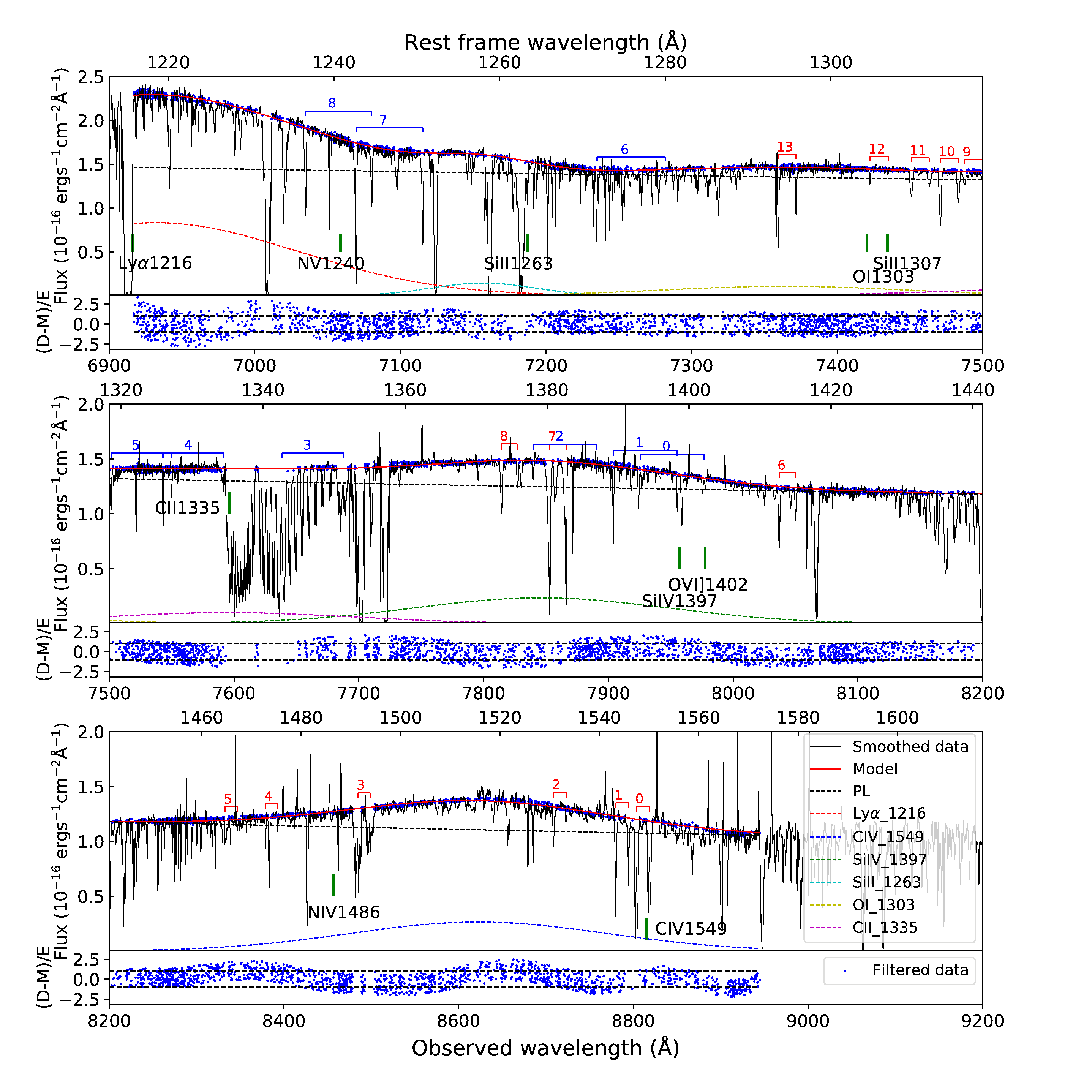}
\caption{Zoom-in spectra of Fig.~\ref{fs} in different wavelength ranges to show the position of various emission and absorption lines. Symbols are the same as in Fig.~\ref{fs}. The color dashed lines represent different model components as denoted in the lower panel. The red and blue bars labeled with sequence numbers above the spectra show the position of \ion{C}{IV} and \ion{Si}{IV} absorption systems which will be further zoomed in Fig. \ref{abs1}.
We also plot the ratio of residual (filtered data - model) to flux error in the lower panel of each sub-figure. 
Two horizontal dashed lines show the positions of $1\sigma$ error level.
\label{em}}
\end{figure*}

The emission features are fitted with the following procedures. We first generate a spline-interpolation function that matches the overall shape of the quasar spectrum. We then filter abrupt emission and absorption features by rejecting data points beyond a tuned flux threshold relative to this spline function.
These abrupt features mainly include sky emission lines, telluric and metal absorption lines, as revealed by gap regions of blue dots shown in Fig. \ref{fs} and \ref{em}. The flux threshold is fixed at 1~$\sigma$ noise level at all wavelengths, which removes most of the abrupt features but still maintains the global shape of the continuum. During these procedures, we did not make judgments in advance on where is the line free region, and what features are real or spurious.
We check the goodness of this filtered featureless spectrum, i.e., the blue dotted line in Fig.~\ref{fs} and \ref{em}, between 6912~\AA~and 8950~\AA~by eye. This wavelength range corresponds to 1215-1573~\AA~in the rest frame, which excludes the complicated Ly$\alpha$ forest and covers the major band of interest (including the \ion{C}{IV} $\lambda\lambda1548,1550$~\AA~and \ion{Si}{IV} $\lambda\lambda1394,1403\rm~$\AA~doublets).

The filtered spectrum are then fitted with a power-law continuum and multiple Gaussian emission lines simultaneously.
We assume every prominent bump above the power law continuum is dominated by emission lines from only one ion. 
For doublets (triplets) ion emission line, we use two (three) Gaussian profiles with their statistical weights set according to \citet[][]{Bal96}. The bumps include the Ly$\alpha$ line at $\sim6920$~\AA, the blueshifted \ion{Si}{IV} lines at $\sim7980$~\AA, the blueshifted \ion{C}{IV} lines at $\sim8610$~\AA, and a few less prominent bumps at $\sim7150-7700$~\AA, all in the observer's frame. 
We attribute the bump feature at $\sim7150$~\AA~to the \ion{Si}{II} $\lambda\lambda\lambda1260,1264,1265$~\AA~lines instead of the \ion{N}{V} $\lambda\lambda1238,1242$~\AA~lines as the latter will be redshifted instead of blueshifted (redshifted by $\sim4100\rm~km~s^{-1}$) which is not expected in a BLR-outflow scenario.
The red peaks of Ly$\alpha$ and \ion{N}{V} emission lines from a rotating accretion disk could contribute to this bump. However, the absence of double-peak features for \ion{C}{IV} and H$\alpha$ emission lines is against this scenario.

The bump near $\sim7400$~\AA~is attributed to the \ion{O}{I} $\lambda\lambda\lambda1302,1304,1306$~\AA~lines instead of the \ion{Si}{II} $ \lambda\lambda1304,1309$~\AA~lines.
The expected positions of the \ion{Si}{II} $\lambda\lambda1304,1309$~\AA~lines are close to the $\sim7400$~\AA~bump. However, the derived parameters of these lines assuming the \ion{Si}{II} line origin is inconsistent with those obtained from the $\sim7150$~\AA~bump (i.e., $v\sim-2660\pm310~\rm km~s^{-1}$ versus $\sim-1320\pm260~\rm km~s^{-1}$ and $FWHM\sim8500\pm400~\rm km~s^{-1}$ versus $\sim3110\pm80~\rm km~s^{-1}$). To test the reliability of identification, we also estimate flux ratios of \ion{O}{I}/\ion{C}{IV} and \ion{Si}{II}/\ion{C}{IV} to be $0.17\pm0.02$ and $0.107\pm0.005$, respectively, which are within the observed ratio range for quasars at $z\sim2-7$ \citep[e.g.,][]{Nagao06,Jua09,De14,Tang19,Onoue20}.
This suggests our identifications of these lines are reliable.

We also add another Gaussian component at $\sim7600$~\AA~and attribute it to \ion{C}{II} $\lambda\lambda\lambda1334,1335,1335$~\AA. Including this component into the global fitting is crucial, although the identification and measurements of this component are difficult due to the significant contamination from the strong telluric features.
In our fitting, absence of this component will lead the model to deviate data at \ion{Si}{II} and \ion{O}{I} regions where emission features are weak. The $\rm \chi^{2}$ reduces from 5450 (absence of this component) to 4878 (including this component). 
The wavelength of this component is fixed to be the expected wavelength of the redshifted \ion{C}{II}, i.e., $v_{\rm CII} = 0 \rm~km~s^{-1}$. This is because that global spline-interpolation function and emission line fitting are difficult to identify this contaminated weak feature, let along measuring its central wavelength. Both freeing its central wavelength or linking it to that of \ion{Si}{II} emission line (similar ionization potential) will significantly increase the $\rm \chi^{2}$ (e.g., 8953 and 5714, respectively).

We emphasize that the above identification of emission lines may be artificial and too ideal. Each Gaussian component could be actually a mixture of different emission lines, which is suggested by the presence of wave-shape residual spectrum as shown in Fig.~\ref{em}. For regions near Ly$\alpha$ and \ion{C}{IV} emission lines, these wave-shape features may be further amplified due to the absence of blue wing data and fringing effects\footnote{In the near-infrared wavelength region, the wood-like response of high-sensitivity scientific CCDs could introduce wave shape features in the spectra.}, respectively.
Besides, fixing $v_{\rm CII} = 0 \rm~km~s^{-1}$ may introduce significant uncertainties on the measurements of \ion{O}{I}, since \ion{O}{I} is weak and close to the \ion{C}{II}.
The \ion{Fe}{II} and \ion{Fe}{III} emission lines \citep{Ves01} and the Balmer continuum are not subtracted in the spectral fitting, as they have relatively small contribution in the band of interest \citep{Shen11}. Including such features in the spectral fitting will introduce extra uncertainties, given the poor long wavelength coverage to constrain them based on the MIKE spectra.

The absorption lines are searched and identified on the red side of Ly$\alpha$ emission line to avoid the contamination of Ly$\alpha$ forests. These procedures are performed by eye,
because telluric and unrelated ion absorption features (e.g., 9th \ion{C}{IV} and 6th \ion{Si}{IV} absorption lines in Fig.~\ref{abs1}) can contaminate these absorption lines, which may lead them to be missed by an automatic-search algorithm.
We first search and identify \ion{C}{IV} absorption features in the column density-redshift space, because \ion{C}{IV} is the most prominent doublets in this wavelength region. 
We calculate the column density by using the apparent optical depth method \citep[][]{Sav91}, with the continuum flux estimated from the spline-interpolation function.
As a \ion{C}{IV} absorber candidate, \ion{C}{IV}$\lambda1548$~\AA~and \ion{C}{IV}$\lambda1550$~\AA~ absorption features should not misalign with each other by $\Delta v \lesssim \Delta v_{FWHM} \sim 10~\rm~km~s^{-1}$, corresponding to $\Delta z\lesssim 0.0001$). 
Besides, we do not require these two lines to have the same column density over the whole line profiles, as they could be potentially contaminated by the telluric and unrelated absorption features.
In the regions where telluric features do not dominate the absorption, we require these two lines to have similar profiles. For those absorption regions dominated by telluric features, the identification is aided by removing these features with the TAPAS telluric template \citep{Ber14} 
\footnote{http://cds-espri.ipsl.upmc.fr/tapas/}.
We notice that TAPAS telluric transmission could not always reproduce the amplitude of absorption features in the MIKE spectrum of BR 1202-0725.
So we prefer to remove the telluric features locally, with the transmission adjusted according to the amplitude of adjacent telluric-dominated absorption features.
We confirm and measure these candidates with the line profile given in equation \ref{efit}.
Once these \ion{C}{IV} absorbers are confirmed, we search for other associated ion absorption features, including all ions identified by emission lines.
Except for those residing on the blue side of the Ly$\alpha$ emission line so highly blended with the Ly$\alpha$ forest, the associated \ion{Si}{IV} absorption features are all identified at the corresponding redshifts.
However, in most cases, we do not detect the corresponding absorption lines from other ions within $250\rm~km~s^{-1}$, which is the width of the broadest absorption \ion{C}{IV} lines identified in the spectrum. 

The form of profile used to confirm and measure the absorbers is as followings: 
\begin{equation}
    \frac{I_{ v}}{I_{\rm 0}} = (1-C_{ v}+C_{ v}e^{-\tau_{\rm 0}\phi_{ v}(v_{\rm c},b)})T(v)^{a_{\rm T}}*G(v),
\label{efit}
\end{equation}
where $I_{v}$ and $I_{0}$ are absorption line and continuum intensities at velocity $v$. $b$, $\tau_{0}$, $v_{\rm c}$ and $\phi_{ v}$ are the Doppler parameter, the optical depth and velocity at the line centre and Voigt profile, respectively. $C_{ v}$ is the covering fraction of the absorbing gas clumps and is in the range of $0 \leq C_{ v} \leq 1$ \citep[e.g.,][]{CC18}. For the narrow absorption lines, 
we assume a constant covering fraction across the line profiles, i.e., $C_{ v}$ = $C_{\rm 0}$.
Since some identified absorption features are significantly contaminated by the telluric features,
we include the correction for these features with the TAPAS telluric transmission template $T(v)$ with
their transmissivity adjusted by a free parameter $a_{\rm T}$. 
In order to fit the narrow absorption lines with 
velocity widths $\lesssim 10 \,\rm km~s^{-1}$ in the MIKE spectra,
we convolve the model profile including telluric features with a Gaussian kernel $G(v)$
that represents the instrumental broadening.
The line spread function of MIKE is obtained by fitting the nearby emission lines in the ThAr lamp spectra.
Our fitting code is based on the python code Lmfit and adopts the Levenberg-Marquardt least-squares method as minimization algorithm \footnote{https://lmfit.github.io/lmfit-py/}. We do not use a uniform quantitative-convergence-criterion for adding components in our fitting, which is usually introduced in the automatically fitting code to stop the fit, such as $(\chi^{2}(\rm new)-\chi^{2}(\rm old))/\chi^{2}(\rm new) < 0.01$ in VPFIT. This is because the significance of a component in our fitting process is not totally accounted by the $\Delta \chi^{2}$, it is also determined by whether it has a corresponding component identified in other ions, since we do not link  \ion{C}{IV} and \ion{Si}{IV} to avoid the artificial definition of a component
(\ion{C}{IV} and \ion{Si}{IV} can have different velocities in such high resolution spectra). We generally stop adding components when the residuals of doublet features are within 1 $\sigma$ noise level.
The best-fit parameters of the emission and absorption lines 
are summarized in Tables~\ref{comp} and \ref{abst}, respectively.

\section{Analyses and Discussions}\label{emlpro}
\subsection{Physical parameters of the SMBH} \label{sec_LM}

For high-$z$ quasars, the mass of the SMBH is usually estimated by using the scaling relations between it and the width of optical or UV emission line.
Assuming the clouds in the BLR are virialized under the gravity of the central SMBH, the BH mass can be estimated as $M_{\rm BH} \sim \frac{\Delta v^{2}r}{G}$, where $\Delta v$ is the velocity dispersion of some broad emission lines and $r$ is the size of the BLR. $\Delta v$ can typically be measured with the \ion{Mg}{II} or \ion{C}{IV} lines, while $r$ can be estimated from the UV continuum luminosity based on the extrapolation of the calibrated radius-luminosity relationship of quasars at low redshifts \citep[e.g.,][]{Kaspi00}.

We estimate the mass of the SMBH in BR~1202-0725 based on the \ion{C}{IV} emission line (emission line properties are summarized in Table~\ref{comp}). Adopting the virial mass estimator developed by \citet[][]{Shen12}, we estimate the SMBH mass to be log$(M_{\rm BH_{CIV}}/M_{\odot})=10.62\pm0.09$. For comparison, \citet[][]{Jun15} obtained the SMBH mass of BR~1202-0725 to be log$(M_{\rm BH_{H\alpha}}/M_{\odot})=10.06\pm0.20$ and log$(M_{\rm BH_{CIV}}/M_{\odot})=10.55\pm0.23$ based on the H$\alpha$ and \ion{C}{IV} lines, respectively. SMBH mass estimated with emission lines from high ions such as \ion{C}{IV} is known to be systematically biased compared to those from low ions such as \ion{Mg}{II} and the hydrogen Balmer lines \citep[e.g.,][]{Shen12,Coat17,Jun17,Ge19,Mar19,Zuo20}.
Such a bias is often attributed to the effects of the non-virialized BLR outflow component \citep[e.g.,][]{Den12}.
Using a sample of 230 quasars at $z = 1.5-4$ with both \ion{C}{IV} and the Balmer lines detected,
\citet{Coat17} quantified such a bias in SMBH mass measurement as a function of the \ion{C}{IV} blueshift from the H$\alpha$ line.
We obtain the SMBH mass to be log$(M_{\rm BH_{CIV,c}}/M_{\odot})=9.65\pm0.10$ after adopting this correction. This value is about 0.4~dex below what is directly measured with the H$\alpha$ line. As the largest \ion{C}{IV}-to-H$\alpha$ blueshift in \citet{Coat17}'s sample is $\lesssim6000\rm~km~s^{-1}$, such an underestimation suggests that the extrapolation of \citet{Coat17}'s relation to a higher BLR outflow velocity (e.g., $\sim 6500 \rm \,km~s^{-1}$ in BR~1202-0725) may no longer be reliable.
In fact, the directly measured correction factor of the \ion{C}{IV} line width based on the detection of the H$\alpha$ line from \citet{Jun15} ($\rm FWHM_{CIV}/FWHM_{H\alpha}\approx2.6$)
is significantly smaller than that ($\approx$ 3.3) 
derived from the extrapolated relation in \citet{Coat17}. This leads to an underestimation in mass by 0.21~dex.

In order to estimate the Eddington ratio, we convert the measured luminosity at rest frame 1350~\AA~to the bolometric luminosity $L_{\rm bol}$ using the empirical relationship from \citet{Shen11}: $L_{\rm bol}=3.81\times\lambda L_{\rm1350~\text{\normalfont\AA}}$.
The estimated the bolometric luminosity of BR 1202-0725 is $L_{\rm bol}=(8.5\pm3.2) \times 10^{47}\, \rm erg~s^{-1}$, and the corresponding Eddington ratios $L_{\rm bol}/L_{\rm Edd}$ are $0.16\pm0.07$ and $1.48\pm0.65$ for the cases without and with
mass correction, respectively.

\subsection{Quasar BLR outflow} \label{sec_outflow}

An accurate determination of the systematic redshift of the quasar is important for the study of its BLR Outflow.
The measured H$\alpha$ \citep{Jun15}, \ion{Mg}{ii} \citep{Iwa02}, CO~$J=1-0$ \citep{{Rie06}} and [\ion{C}{II}] 158$\mu \rm m$ \citep{Wagg12} redshifts for BR 1202-0725 are $4.689\pm0.005$, 4.689, $4.6949\pm0.0003$,  and 4.6943, respectively. We herein adopt the H$\alpha$ emission line redshift from \citet{Jun15} as the systematic redshift of BR 1202-0725. This is because Ly$\alpha$ emission line in our MIKE spectrum is highly contaminated by the strong absorption features on its blue wing, and most other prominent emission line features are produced by high ions which may be largely contributed by the BLR outflow. Although far-infrared (e.g., [\ion{C}{II}] 158$\mu \rm m$) and millimeter (e.g., CO~$J=1-0$) emission lines may provide better redshift estimates, as these lines likely emitted by the cold interstellar medium of the host galaxy, 
whose redshift could represent that of SMBH.
We still prefer to use H$\alpha$ emission line as the systematic redshift to make the comparisons of emission line dynamics among different works easier. Compared to far-infrared and millimeter lines, H$\alpha$ emission lines are more often observed and used in the dynamical study of BLR outflow and their impacts on C IV single-epoch virial mass estimates \citep{Plo15,Coat17,Jun17}.
We notice that the difference in the derived BLR outflow velocities introduced by different choices of systematic redshift does not affect our main conclusions.  
The expected positions of some strong emission lines at the systematic redshift are labeled in Fig.~\ref{fs} and Fig.~\ref{em}. Most of the emission lines from high ions are clearly blueshifted (Table~\ref{em}). 
For example, the measured redshift of the \ion{C}{IV} and \ion{Si}{IV} lines are $4.5665\pm0.0005$ and $4.611\pm0.002$, respectively, corresponding to outflow velocities of $-6460\pm260$ and $-4090\pm270 \rm~km~s^{-1}$.

A highly ionized quasar outflow seems to be further confirmed by the detection of a \ion{C}{IV} and \ion{Si}{IV} absorption system with blueshifted velocity of $\approx -1.1\times10^{4}\rm~km~s^{-1}$ and velocity width of $\approx 250 \rm~km~s^{-1}$, which reside on the blue wings of the corresponding blueshifted broad emission lines (i.e., the 3rd absorption system in Fig.~\ref{em}).
Such P-Cygni profile is possibly produced when the cold gas clumps entrained in the foreground outflow shield the emission from outflow in the background. The narrow outflow absorbers (e.g, $FWHM \lesssim 700 \rm ~km~s^{-1}$ in the intermediate-resolution spectra) are found to be common for quasars in the statistic studies \citep{Wey79,Nes08,Wild08,Per18,CC20}. For example, by decomposing the observed $\sim 25000$ narrow \ion{C}{IV} absorbers within BOSS DR12 quasar spectra into several populations including outflows, \citet{CC20} found that the majority of absorbers with blueshift velocities $\sim 1000-8000 \rm ~km~s^{-1}$ are from quasar outflows. The velocity of these narrow outflow absorbers could be up to $\sim -1.4 \times 10^{4} \rm~km~s^{-1}$ \citep[e.g.,][]{Wild08,Ham11}. 

These narrow outflow absorbers are important for the structure and dynamic studies of outflows.
Although broad emission lines are usually used to reveal the existence of BLR outflows,
it is difficult to study the properties of outflow such as density, temperature and ionization parameters, which is crucial to understand their physical origins and influences on their host galaxies.
This is because emission lines from different ion BLR outflows are likely dominated by the gas in different spacial and dynamic regions, which is suggested by the significant differences in emission line dynamic properties as summarized in Table \ref{comp}.
Nevertheless, the matched line velocities and profiles, as revealed in Fig.~\ref{abs1}, indicates that line ratios of these narrow absorption features are still good tracers of outflow physical parameters. Absorption lines have been used to study  the structure and dynamic studies of outflow \citep[e.g.,][]{Ham11,CC18,He19,Zhao21}.  

To study the structure and dynamics of the outflow within BR 1202-0725, we use a few criteria in the literature \citep[i.e., partial covering, line-locking signature, smooth profile across broad line,][]{Bar97,CC19} to examine the origin of identified 14 narrow \ion{C}{IV} absorption systems (see a summary of these systems in Table~\ref{abst} and zoom-in spectra of them in Fig.~\ref{abs1}). Except for component ``c'' of the 0th absorption system (partial covering, broad and close to quasar), as listed in Table~\ref{abst} and shown in Fig.~\ref{abs1}, we can not rule out the possibility that these absorption features are from the intervening inter-galactic absorbers \citep[e.g.,][]{Has20}. The derived velocity for component ``c'' of the 0th absorption system is small and has a large error (i.e., $-124.5\pm264~\rm km~s^{-1}$), therefore it may be produced by any kinds of gas closed to quasar (e.g., outflow gas, infalling gas and gas within host galaxies). We emphasize that although the partial covering nature of this component is not directly constrained by data points, the existence of partial covering is still indicated by the $\chi^{2}$ tests. The $\chi^{2}$ of model fitting with this partial covering component is 724, while this value changes to $870$, $773$ and $750$ when this component is absent, is replaced by one and two fully covering components, respectively.
Since most of the identified \ion{C}{IV} absorbers could be the intervening inter-galactic absorbers, we leave further analyses and discussions of these narrow absorbers in a separate paper.

\subsection{Comparison to other quasar samples}\label{subsec:Comparison}

In this section, we compare BR 1202-0725 to some other samples of quasars showing outflows. In Fig.~\ref{vew}, we compare the \ion{C}{IV} emission line based outflow velocity of BR 1202-0725 to the SDSS quasars from \citet[][]{Shen11}. As the redshift and outflow velocity of the SDSS quasars are limited by the wavelength coverage of the spectrograph, we also plot some other quasar samples with high reported \ion{C}{IV} blueshift velocities at various redshifts for comparison, including those from \citet{Wu11,Luo15,Plo15,Jun17,Sul17,Vietri18,Meyer19,Sch20}. 
As shown in this figure, the outflow velocity of BR 1202-0725 traced by the \ion{C}{IV} emission line is about the highest at $z>4$.

We also compare the properties of emission lines from different ions in the outflow. 
As shown in the upper panel of Fig.~\ref{ipfwhm}, the emission lines from the lower ions in BR 1202-0725, e.g., \ion{Si}{II}, are significantly narrower and less blueshifted compared to those from the higher ions such as \ion{C}{IV}. 
This trend is commonly observed in active galactic nuclei with broad emission lines \citep[e.g.,][]{Netzer13,Onoue20}, which suggests that the outflow is mostly photo-ionized by the UV photons from the accretion disk and is stratified, with the higher ions distributed in the inner layers so have higher velocities and larger velocity dispersion. 

We further compare the above trend from quasars at different redshifts in the lower panel of Fig.~\ref{ipfwhm}, where 
we obtain the mean velocities of emission lines from different ions of a few quasars samples across $z = 1.5-7.5$ from \citet{Meyer19}.
Meyer et al. (2019) included several sets of data collected in the literature. We divide these data into three quasar samples : SDSS survey data (DR7 and DR12) at $1.5<z<4.5$, XQ100 and Giant Gemini GMOS Survey data at $3.5<z<5.5$ and quasars at $5.4<z<7.5$ they collected in the literature. Since the number of SDSS quasar ($\sim100000$) is much larger than other surveys ($\sim100$), we prefer to leave the SDSS data alone and combine the XQ100 and Giant Gemini GMOS Survey as quasars at $3.5<z<5.5$.
All blueshift velocities in \citet{Meyer19} are measured relative to the \ion{Mg}{II} emission line, and we have excluded quasars showing redshifted \ion{C}{IV} and \ion{Si}{IV} lines in order to avoid the contamination from any inflow features.
The general trend that higher ions tend to have larger blueshifts is clear at all redshifts, although the trend may be slightly more significant at the highest redshift bin ($5.4<z<7.5$). This stronger trend at the highest redshift is not an artifact caused by luminosity bias, as the samples used in their study are luminosity matched.
We notice that emission lines from BR 1202-0725 shows significantly larger blueshifts and steeper gradient of blueshift with respect to the ionization potential than the mean values of quasars at all redshifts. This probably indicates this quasar hosts an extraordinarily fast BLR outflow. The fast BLR outflow within BR 1202-0725 are likely not caused by the additional powers provided by an adjacent strong jet, as BR 1202-0725 is a radio quiet quasar \citep[i.e., $R\equiv f_{5 \rm~GHz}/f_{4400~\rm \text{\normalfont\AA}} <1.3$,][]{Vig05}. Luminosity ($L_{\rm bol}\sim10^{48}\rm~erg~s^{-1}$) and accretion rate ($L_{\rm bol}/L_{\rm Edd,H\alpha} \sim 0.6$) may not be the main physical origins for such BLR outflow as well. A large fraction of quasars with comparable luminosity (as shown in Fig. \ref{vew}) and accretion rate show much slower BLR outflow. The extraordinarily fast BLR outflow observed in BR 1202-0725 may just be a result of orientation effect: a lucky alignment of the line-of-sight with the main body of the non-spherical BLR outflow. Such orientation-dependent BLR outflow scenario was proposed to explain the distribution of the observed (non-variable) C IV emission profiles \citep[][]{Den12}.

The \ion{O}{I} and Ly$\alpha$ emission lines from BR 1202-0725 seem to depart from the trends shown in Fig.~\ref{ipfwhm}. They are significantly broader than expected from a low ionization line in an outflow. Furthermore, the \ion{O}{I} also appears to be blueshifted to an unusually large velocity. This discrepancy could partially be attributed to the origin of these low ionization lines, which may not due to the BLR outflows, for example, virialized BLR gas. 
The large uncertainties in the measurements of Ly$\alpha$ and \ion{O}{I} may also play roles on the discrepancy. For Ly$\alpha$ line, the absence of the blue wings makes it difficult to determine the peak of emission and to subtract the \ion{N}{V} contribution. Therefore, the measured velocity and $FWHM$ may not reflect the true dynamics of Ly$\alpha$ emission lines. For \ion{O}{I} emission line, \ion{Si}{II} $\lambda\lambda1304,1309$~\AA~emission is not the main contamination. Adding the \ion{Si}{II} $\lambda\lambda1304,1309$~\AA~does not change the fitting result significantly, if the velocity and line width of this doublet are linked to \ion{Si}{II} $\lambda\lambda1260, 1264, 1265$~\AA~and normalization is set to be free (not the same excited state). If the normalization is also linked, this bump will be broken by \ion{Si}{II} $\lambda\lambda1304,1309$~\AA~into two parts of emission features. There is no way to fit the remain part of the bump with the emission from just one ion such as \ion{O}{I}, which means a more complicated and fine-tuned model is needed. The weak and broad \ion{O}{I} may be instead affected by the adjacent \ion{C}{II} emission lines, whose measurements are uncertain due to the contamination of telluric features. 

\section{Summary} \label{sec_summary}

We report a strong BLR outflow from a $z=4.689$ quasar BR~1202-0725 based on the high signal-to-noise ($\rm S/N\sim 38$ per pixel), high-resolution ($R\sim32000$) spectrum taken with the MIKE spectrograph on the 6.5m Magellan/Clay telescope. 

The MIKE spectrum of BR~1202-0725 is characterized by a lot of emission lines in a wide velocity range blueshifted from the H$\alpha$ emission, indicating there is a strong BLR outflow. The highest blueshifted emission lines are produced by high ions. The most significant ones are the \ion{C}{IV} $\lambda\lambda1548,1550$~\AA~and \ion{Si}{IV} $\lambda\lambda1394,1403$~\AA~doublets, which have been blueshifted to $\sim-6500$ and $-4100\rm~km~s^{-1}$. We have also discovered 14 narrow \ion{C}{IV} absorption systems. Most of them are likely from the intervening intergalactic medium, but some multi-component absorption lines are possibly associated with the outflow and have been further blueshifted up to $\approx -1.1 \times 10^{4}\rm~km~s^{-1}$. 

We further compare BR~1202-0725 to other quasar samples. We find that its BLR outflow velocity traced by the UV emission lines is among the highest ones in quasars at $z>4$. The measured properties of UV emission lines from different ions also follow a clear trend that higher ions tend to be broader and outflow at higher average velocities, a common trend also detected in other quasars. This trend suggests the BLR outflow is photo-ionized by the UV radiation from accretion disk and is stratified, with the higher ions distributed in the regions closer to SMBHs, so they have higher velocities and larger velocity dispersion. Some low ionization emission lines, such as \ion{O}{I} and Ly$\alpha$, do not follow this trend defined by emission lines from high ions. They are probably from the virialized BLR gas instead of the BLR outflow, or they could be highly contaminated by other spectral features in the adjacent wavelength range.

\begin{figure*}
\centering
\includegraphics[width = .9\linewidth]{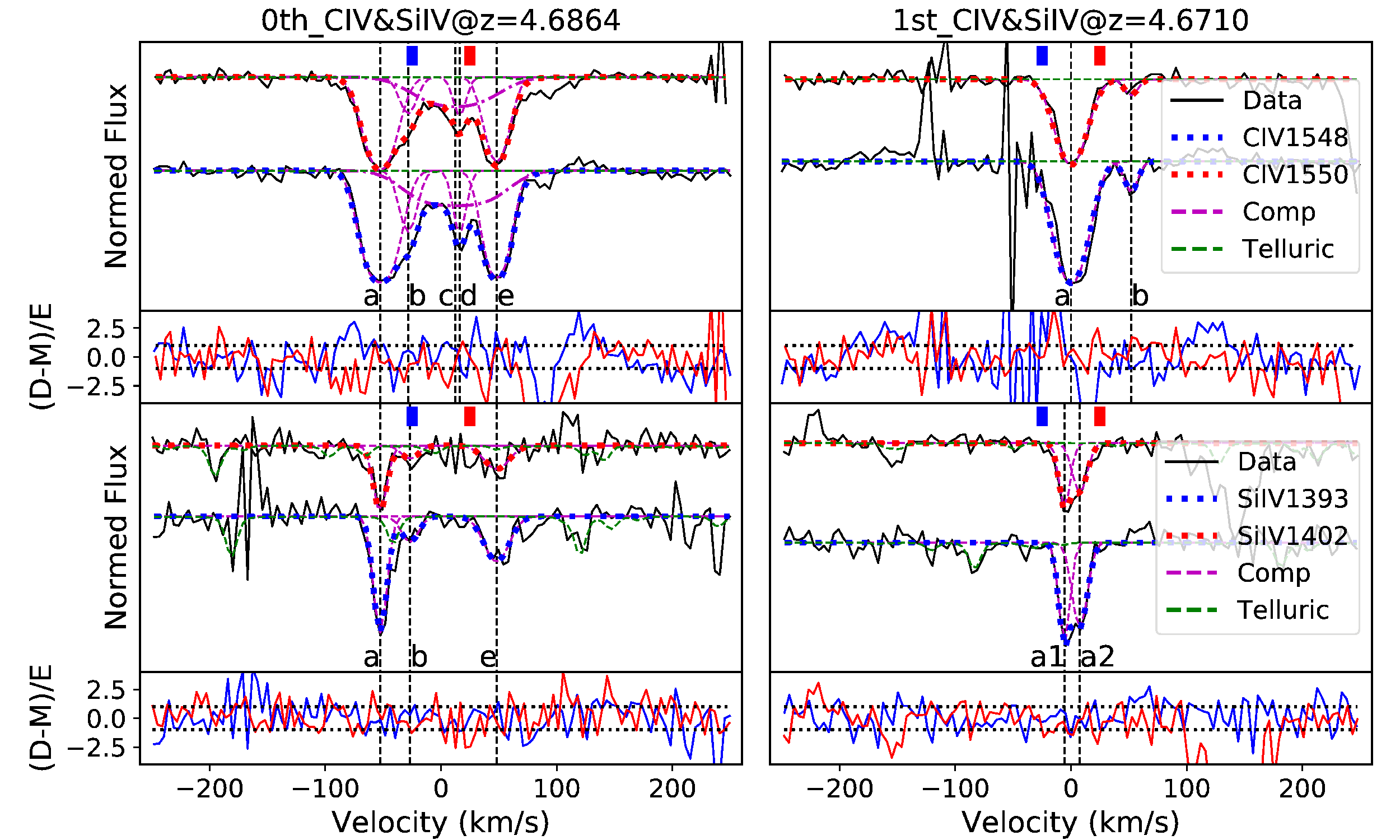}
\includegraphics[width = .9\linewidth]{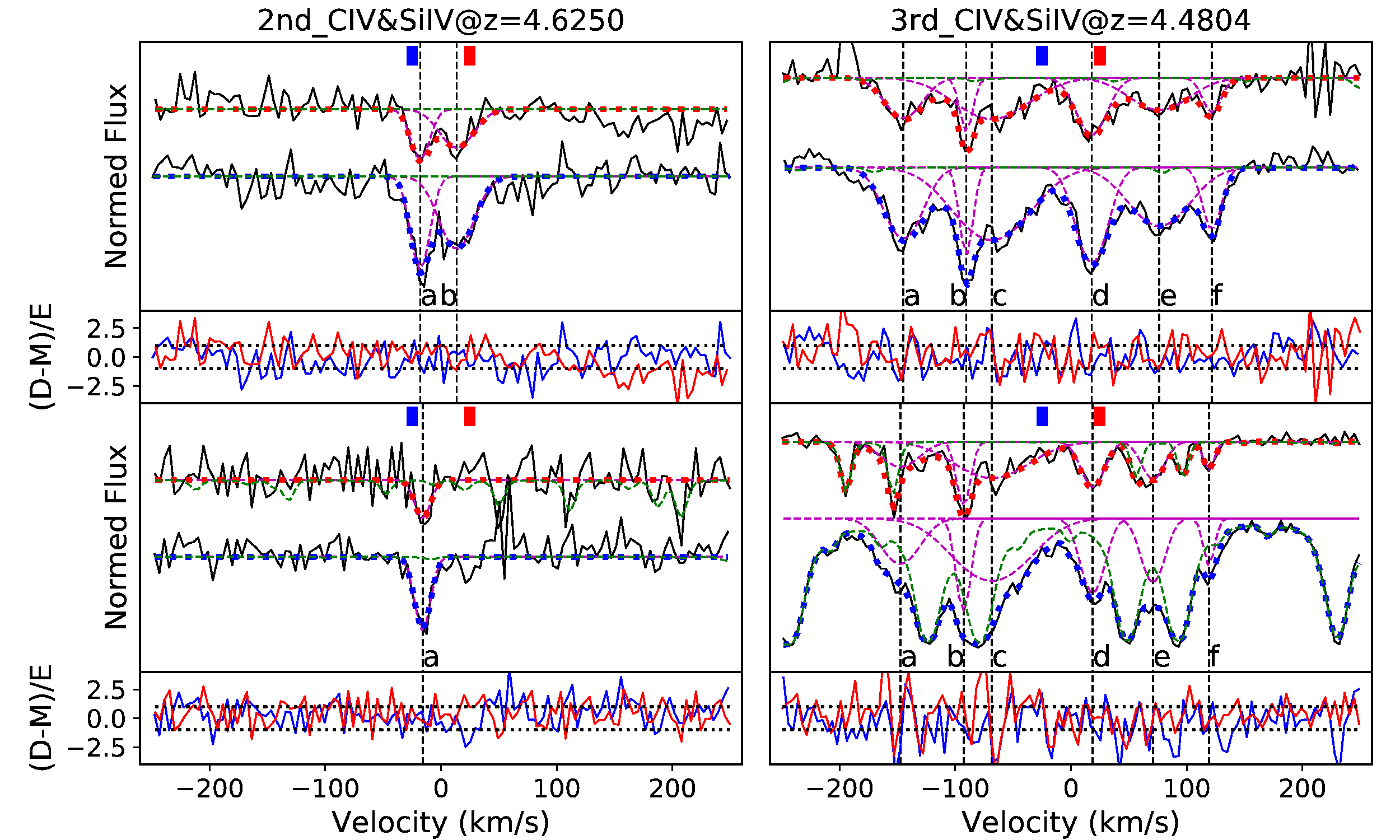}
\caption{
The identified \ion{C}{IV} and \ion{Si}{IV} absorption lines in BR 1202-0725. The positions of these absorption lines are coded with numbers shown in Fig.~\ref{em}. Different sub-figures are ranked according to their redshifts as labeled on the top of each sub-figure.
In the upper panel of each sub-figure, the black curves represent the MIKE data. The red and blue dotted curves generally represent models of the doublets, except for the 3rd absorption features, we use them to represent models including telluric absorption features. Except for that dot-dashed lines represent ``c'' component of the 0th \ion{C}{IV} absorption system, magenta dashed lines are used to represent individual model components, which are labeled with letters as summarized in Table \ref{abst}. Green dashed lines represent the fitted TAPAS telluric templates. The widths of the colored bars show the FWHM (in $\rm km~s^{-1}$) of MIKE at the corresponding wavelength. In the lower panel of each sub-figure, colored solid curves show the residual of the corresponding model (including telluric features), e.g., (Data-Model)/Error. The horizontal dotted lines represent the positions of 1$\sigma$ error level.}
\label{abs1}
\end{figure*}

\addtocounter{figure}{-1}
\begin{figure*}
\centering
\includegraphics[width = 1.\linewidth]{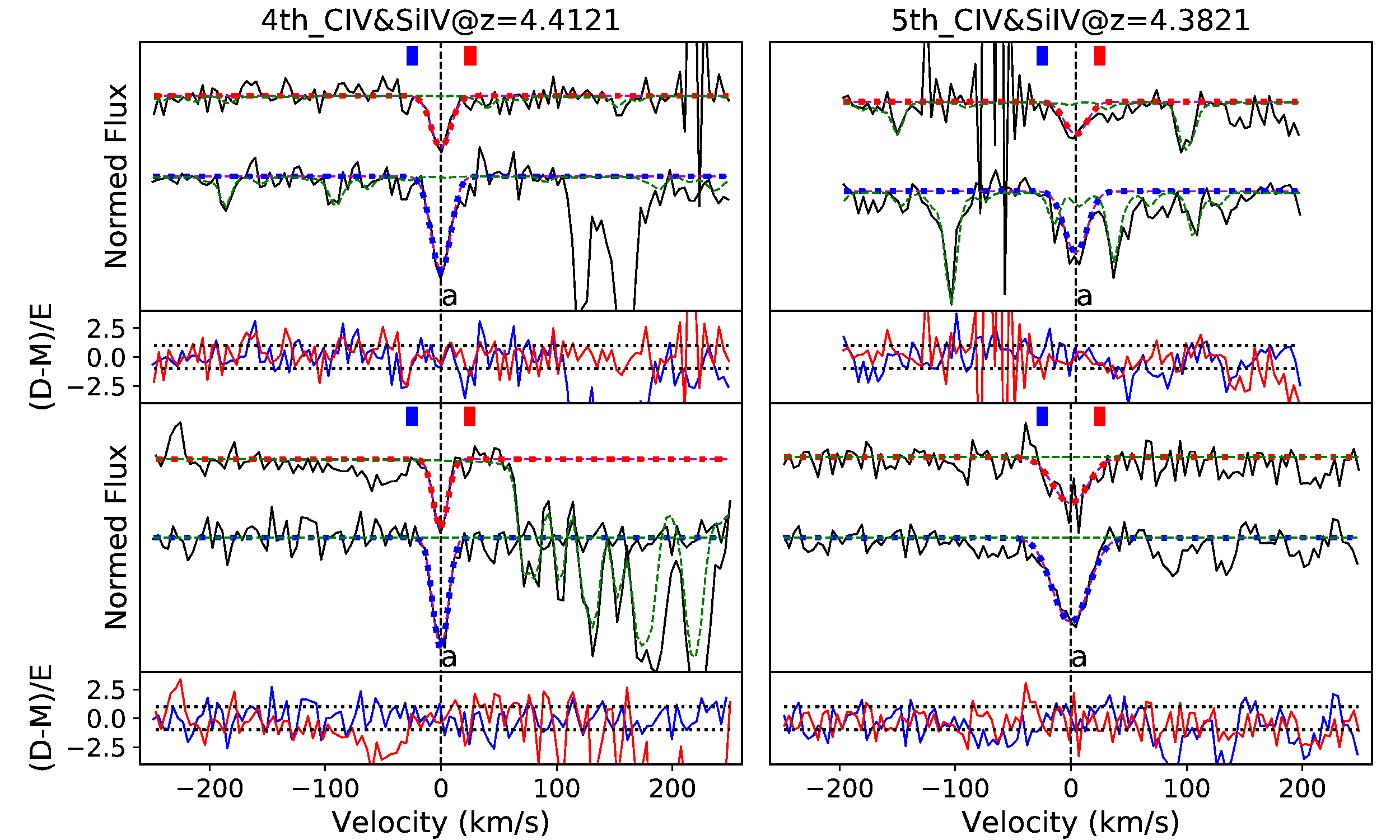}
\includegraphics[width = 1.\linewidth]{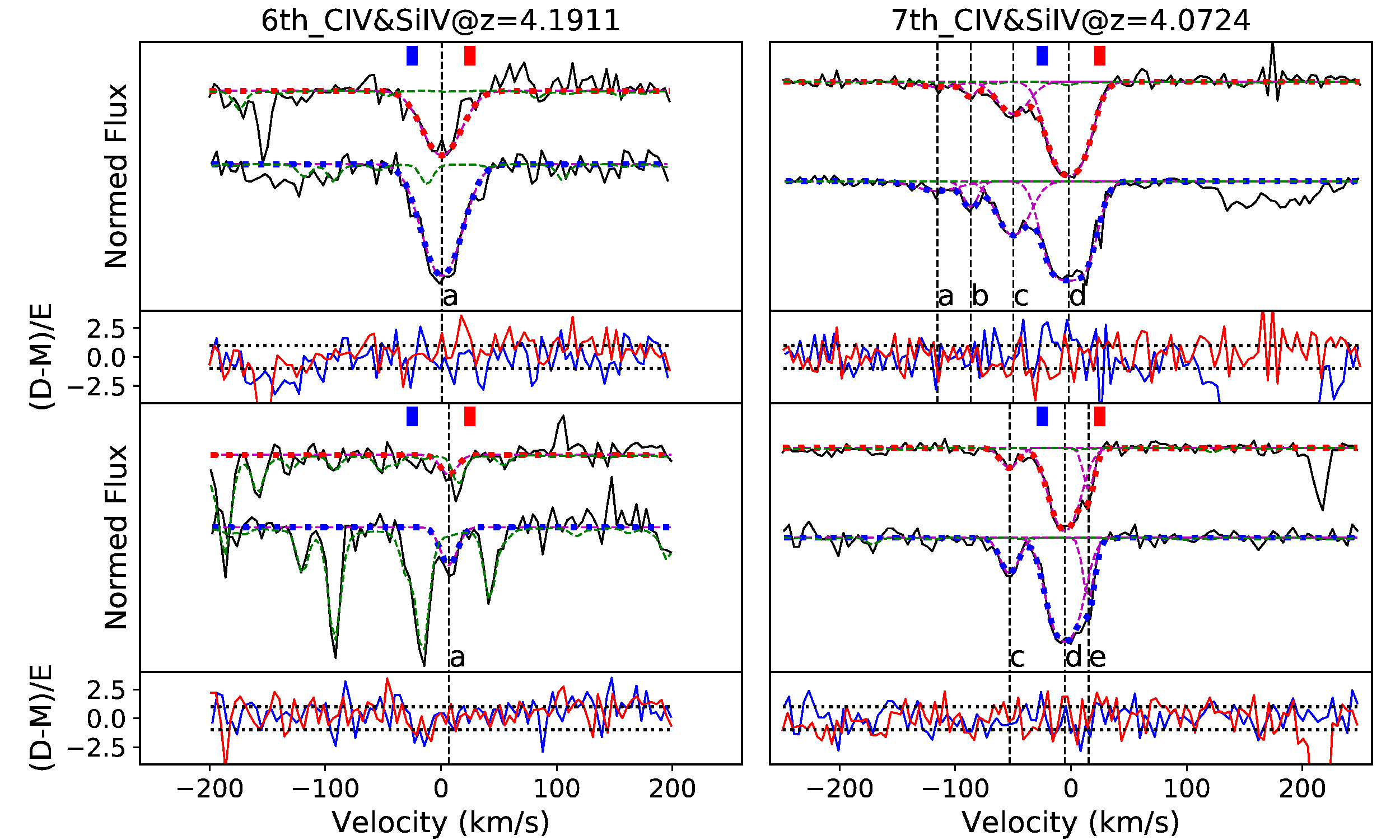}
\caption{Continued}
\label{abs2}
\end{figure*}

\addtocounter{figure}{-1}
\begin{figure*}
\centering
\includegraphics[width = 1.\linewidth]{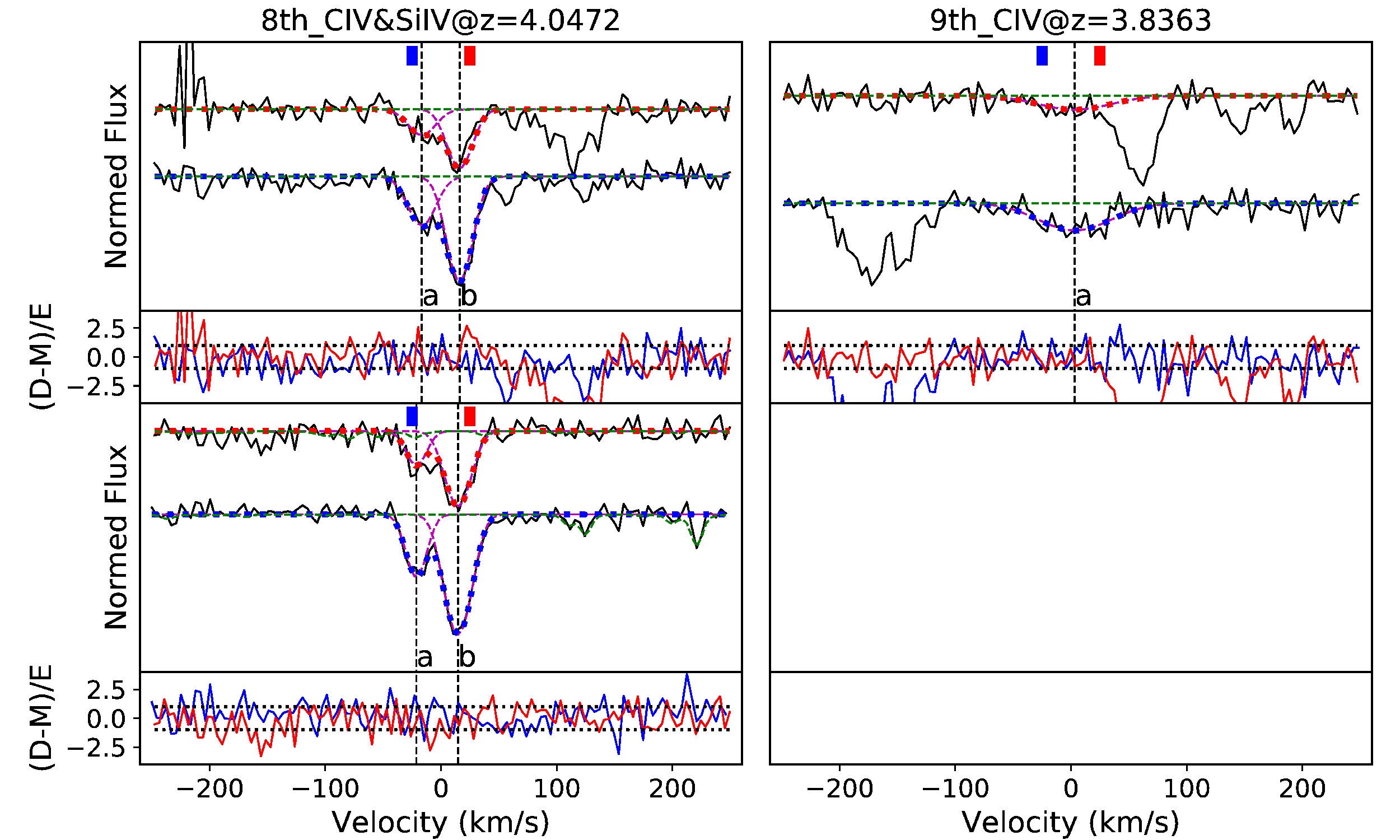}
\includegraphics[width = 1.\linewidth]{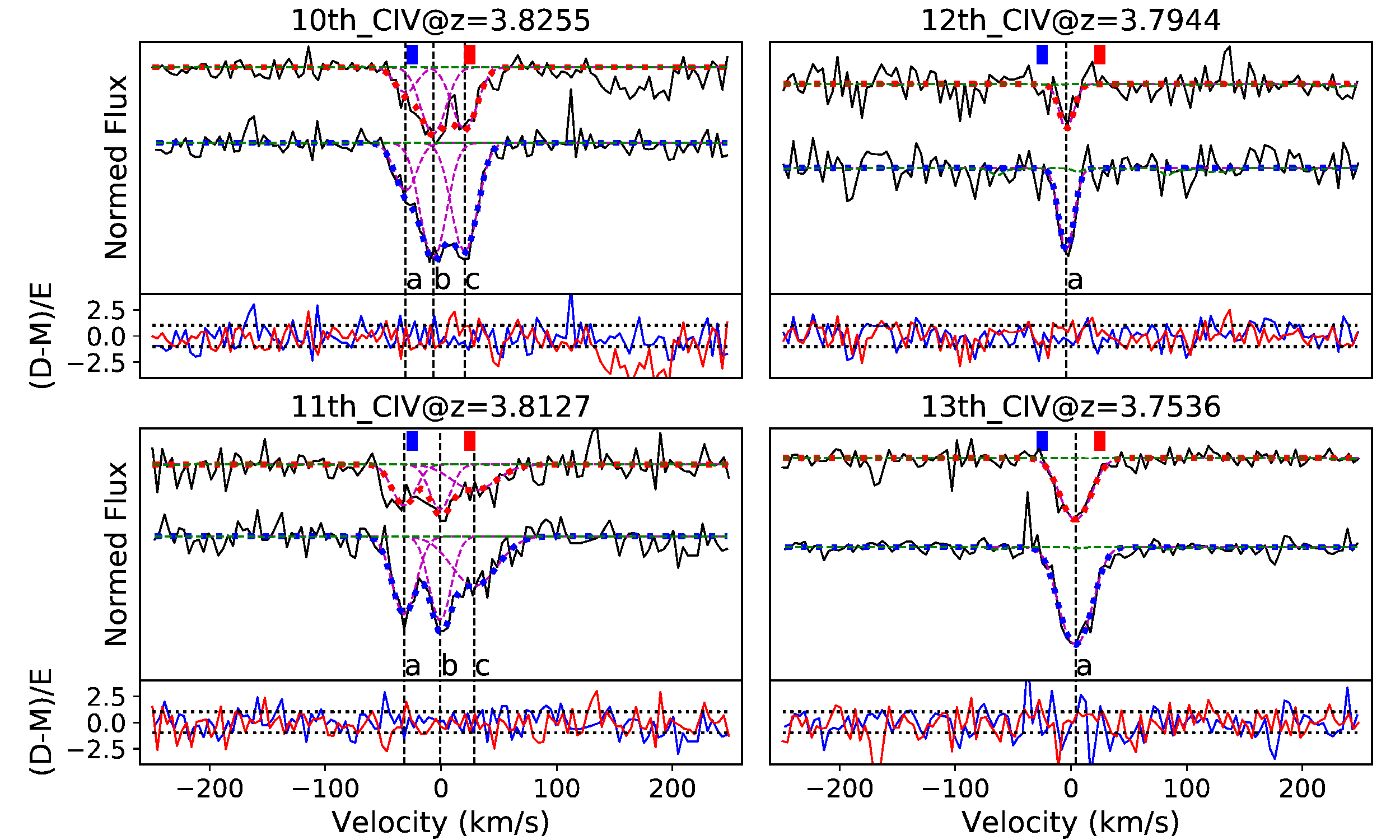}
\caption{Continued}
\label{abs3}
\end{figure*}

\begin{figure*}
\centering
\includegraphics[width = 1\linewidth]{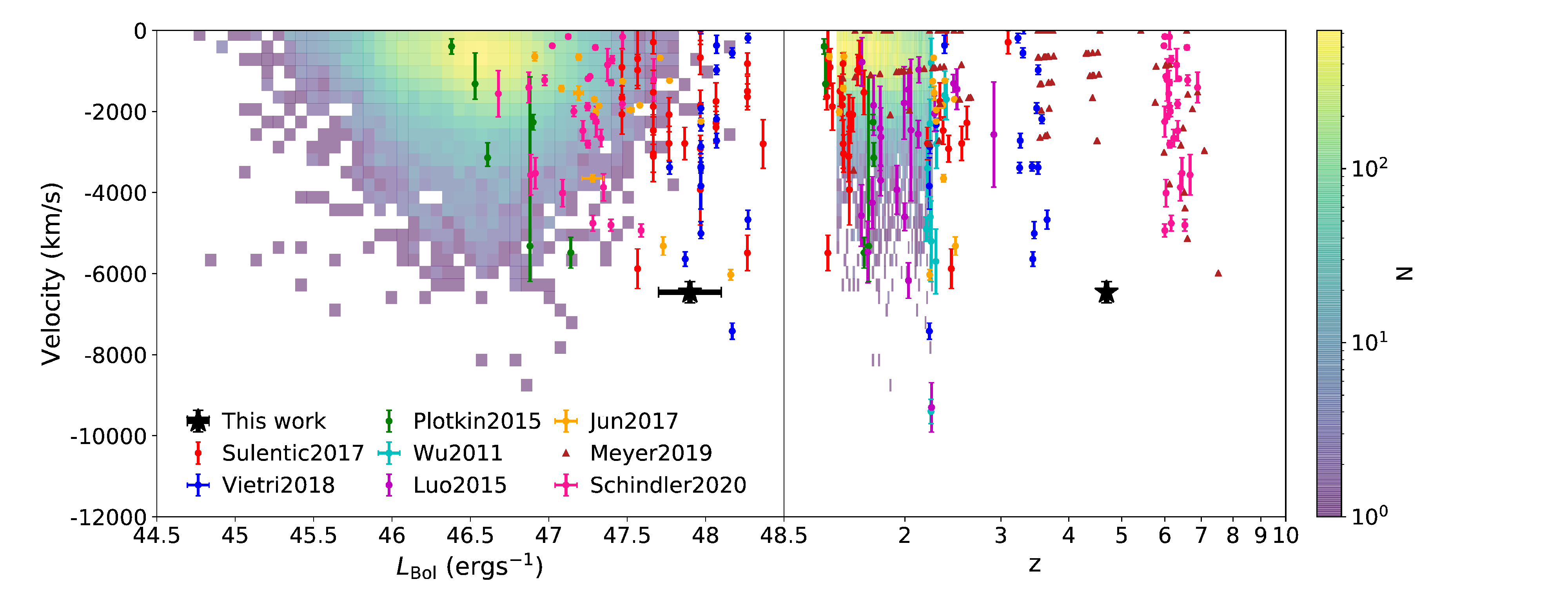}
\caption{Comparison of the measured blueshift velocities of \ion{C}{IV} emission lines for different quasar samples. Left: BLR outflow velocity versus bolometric luminosity. Right: BLR outflow velocity versus redshift. The plotted quasar samples include the SDSS quasars \citep[][2D density map with colorbars denoting the number of quasars]{Shen11} and some other quasar samples with high \ion{C}{IV} blueshift velocity reported in the literature \citep{Wu11,Luo15,Plo15,Jun17,Sul17,Vietri18,Meyer19,Sch20}. The clear velocity gaps in \citet{Meyer19}'s quasar sample are results of rounding issue. Redshift references: H$\alpha$ emission line for this work and \citet[]{Plo15,Jun17}, H$\beta$ emission line for \citet[]{Wu11,Luo15,Plo15,Sul17,Vietri18}, [\ion{O}{III}] emission line for \citet[]{Plo15}, \ion{Mg}{II} emission line for \citet[]{Wu11,Luo15,Meyer19} and SDSS sample,  [\ion{C}{II}] emission line for \citet[]{Sch20}, SDSS CAS for \citet[]{Wu11,Luo15}, absorption system for \citet[]{Wu11,Luo15}. Note: The bolometric luminosity of quasars in this figure is not directly referenced from the literature. Instead, all bolometric luminosity is calculated from $L_{5100\rm~\text{\normalfont\AA}}$ with bolometric correction set to be 9.26 \citep{Shen11} for consistency.}
\label{vew}
\end{figure*}

\begin{figure*}
\centering
\includegraphics[width = 1.\linewidth]{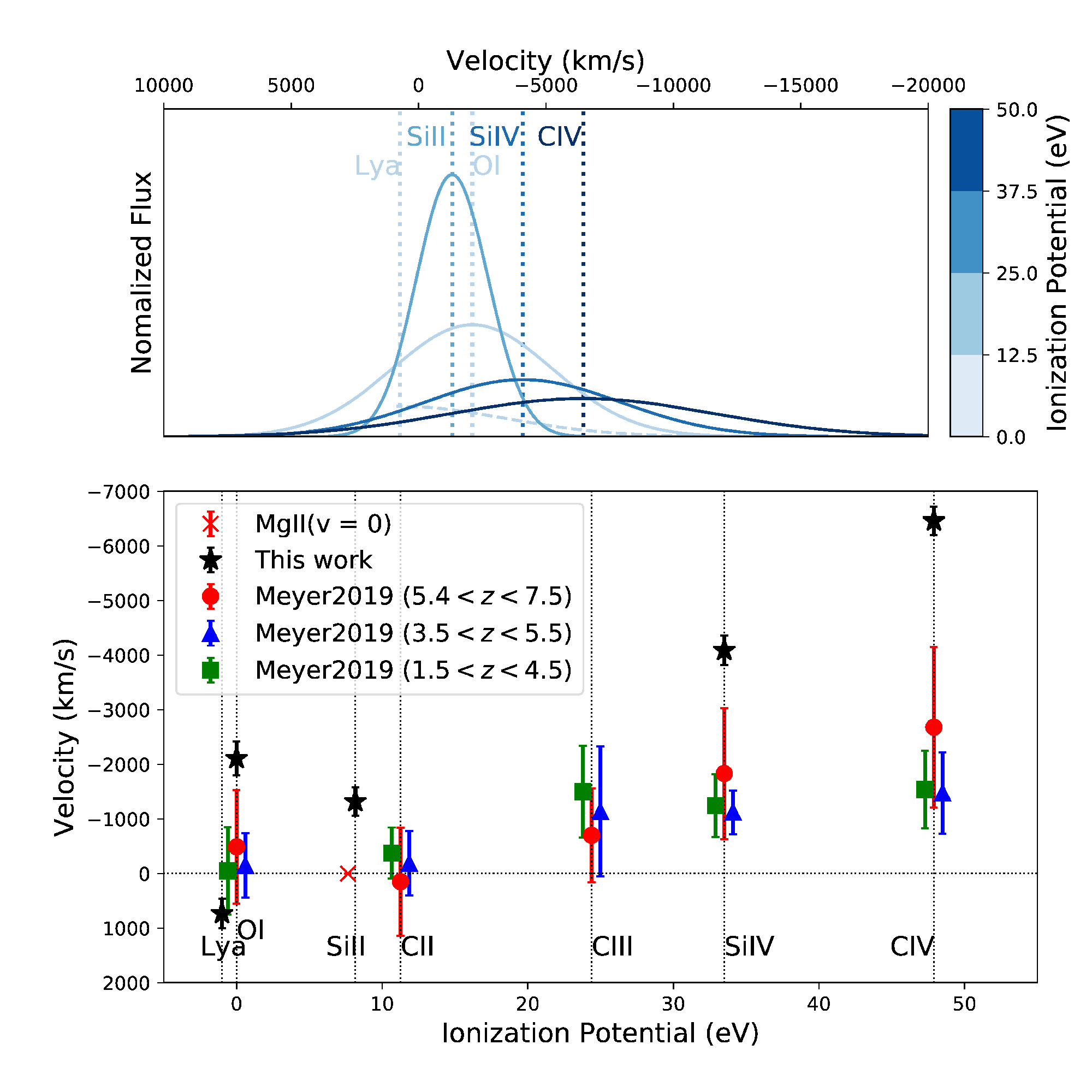}
\caption{
Upper panel: comparison of the best-fit profiles of the detected broad emission lines in velocity space. Dashed and solid lines represent hydrogen Ly$\alpha$ and metal emission lines, respectively. The color vertical dotted lines represent the peak velocities of corresponding lines. Line strengths are scaled so that the integrated fluxes of different lines are the same.
Lower panel: blueshift velocities of the emission lines versus ionization potential of different ions. Black stars represent measurements in this work. The quasars in \citet{Meyer19} include (1) SDSS survey at $1.5<z<4.5$, (2) XQ100 and Giant Gemini GMOS Survey at $3.5<z<5.5$, and (3) quasars at $5.4<z<7.5$ they collected in the literature.
The blueshifts of \citet{Meyer19} are measured relative to the \ion{Mg}{II} emission line. Error bars of their data show the standard deviations of the emission line velocities at each redshift bin. Ionization potentials of lines are shifted for clarification. To avoid contamination from inflow features, we exclude sources in \citet{Meyer19} with redshifted \ion{C}{IV} or \ion{Si}{IV} emission lines.
Note that the redshift of the \ion{Mg}{II} emission line of BR 1202-0725 \citep{Iwa02} is the same as that of the H$\alpha$ emission line \citep{Jun15}.}
\label{ipfwhm}
\end{figure*}

\begin{table*}
\caption{Properties and MIKE observations of BR 1202-0725. $^{\rm a}$ From \citet{Jun15}. $^{\rm b}$ From \citet{Cutri03}. $^{\rm c}$ From \citet{Vig05}. $^{\rm d}$ Denote the parameters corrected for outflow effects using \citet[][]{Coat17}'s relation.}
\label{obs}
\begin{center}
\begin{tabular}{cl}
\hline
Properties of BR 1202-0725\\
\hline
R.A. (J2000)			& 12:05:23.12\\
Dec. (J2000) 			& -07:42:32.11 \\
Redshift$^{\rm a}$ & 4.689$\pm$0.005    \\
$m_{\rm J}$/mag$^{\rm b}$			& 16.759 \\
$m_{\rm H}$/mag$^{\rm b}$			& 15.781 \\
$m_{\rm K}$/mag$^{\rm b}$			& 15.544 \\
$f_{\rm0.5-2 keV}$ ($\rm erg~s^{-1}~cm^{-2}$)$^{\rm c}$ & $\rm 8.5\times10^{-14}$ \\
$\alpha_{\rm ox}$$^{\rm c}$ & -1.8                      \\
$R=f_{5 \rm~GHz}/f_{4400~\rm \text{\normalfont\AA}}$$^{\rm c}$        & $<1.3$         \\
\hline
Magellan/MIKE Observations\\
\hline
Date                &   2019-04-27  \\
Airmass             &   1.39        \\
Seeing              &   $0\farcs6$        \\
Slit width          &   $0\farcs7\times5\farcs0$ \\
Exposure time (s)   &   $6 \times 1800$\\
$\rm S/N_{@7500,8000,8600~\text{\normalfont\AA}}$ & 35,36,40  \\
\hline
Derived parameters\\
\hline
$L_{\rm bol}$ ($10^{47}\rm~erg~s^{-1}$)     & $8.5\pm3.2$      \\
log$(M_{\rm BH(C IV)}/M_{\odot})$  & $10.62\pm0.09 $  \\
log$(M_{\rm BH(C IV,c)}/M_{\odot})$$^{\rm d}$ & $9.65\pm0.10$ \\
$L_{\rm bol}/L_{\rm Edd}$ 		        & $0.16\pm0.07$   \\
$L_{\rm bol}/L_{\rm Edd,c}$$^{\rm d}$		& $1.48\pm0.65$   \\
\hline
\label{tab:simulations-overview}
\end{tabular}
\end{center}
\end{table*}

\begin{table*}
\caption{Continuum and Emission Line Properties of BR 1202-0725. $^{\rm a}$ H$\alpha$ redshift measured in \citet{Jun15} is referred as systematic redshift of BR 1202-0725. The negative and positive velocities represent blueshift and redshift relative to this systematic redshift. The definition of the Ly$\alpha$ redshift is the peak of the Gaussian profile, which is obtained by fitting the red wing of Ly$\alpha$ emission line. $^{\rm b}$ REW : rest-frame equivalent width. $^{\rm c}$ The identification and measurement of \ion{C}{II} are highly uncertain, due to the significant telluric contamination. Lack of such component will lead to bad fit at wavelength region from 7100 to 7600~\text{\normalfont\AA}. We fix the velocity of this emission component to be 0$~\rm km~s^{-1}$.}
\begin{center}
\label{comp}
\begin{tabular}{cl}
\hline
Power-law index         & $-1.41\pm 0.03$      \\
$\lambda L_{\rm 1350~\text{\normalfont\AA}} (10^{47} \rm erg~s^{-1}$) 	& $2.2\pm 0.8 $	\\
$v_{\rm Ly\alpha}$($10^{3}\,\rm km~s^{-1}$)$^{\rm a}$   & $0.73 \pm 0.27$   \\
REW$\rm _{Ly\alpha}$ (\AA)$^{\rm b}$ 		             & $23 \pm 9	$	\\
$\rm FWHM_{\rm Ly\alpha}$ ($10^{3} \,\rm km~s^{-1}$)    & $9.70 \pm 0.09$ \\
$v_{\rm C IV}$ ($10^{3}\,\rm km~s^{-1}$)		         & $-6.46 \pm 0.26$	\\
REW$\rm _{C IV}$ (\AA) 			             & $17 \pm 6 $		\\
$\rm FWHM_{C IV}$ ($10^{3}\,\rm km~s^{-1}$)		     & $12.4 \pm 0.1$	\\
$v_{\rm Si IV}$ ($10^{3}\,\rm km~s^{-1}$)		         & $-4.09 \pm 0.27 $	\\
REW$\rm _{Si IV}$ (\AA) 			             & $ 8.1 \pm 3.1$	\\
$\rm FWHM_{\rm Si IV}$ ($10^{3}\,\rm km~s^{-1}$)	     & $8.88  \pm 0.14$	\\
$v_{\rm Si II}$ ($10^{3}\,\rm km~s^{-1}$)		 & $-1.32 \pm 0.26 $\\
REW$\rm _{\rm Si II}$ (\AA) 			     & $1.4 \pm 0.5$	\\
$\rm FWHM_{\rm Si II}$ ($10^{3}\,\rm km~s^{-1}$)	 & $3.11  \pm 0.08$	\\
$v_{\rm O I}$ ($10^{3}\,\rm km~s^{-1}$)		 & $-2.11 \pm 0.31 $\\
REW$\rm _{\rm O I}$ (\AA) 			         & $2.3 \pm 0.9$	\\
$\rm FWHM_{\rm O I}$ ($10^{3}\,\rm km~s^{-1}$)	 & $7.4\pm0.4$	\\
$v_{\rm C II}$ ($10^{3}\,\rm km~s^{-1}$)$^{\rm c}$		 & 0 (fixed)\\
REW$\rm _{ C II}$ (\AA) 			     & $3.0 \pm 1.1$	\\
$\rm FWHM_{\rm C II}$ ($10^{3}\,\rm km~s^{-1}$)	 & $9.3\pm1.1$	\\
\hline
\label{comp}
\end{tabular}
\end{center}
\end{table*}

\begin{table*}
\caption{Identified \ion{C}{IV} and \ion{Si}{IV} absorption systems of BR 1202-0725. For absorption features at $z < 4.0$, the \ion{Si}{IV} lines cannot be well fitted because of the contamination from the Ly$\alpha$ forest. Systematic uncertainties related to the continuum fitting are not included in the errors of this table.}
\begin{small}
\label{abst}
\begin{center}
\begin{tabular}{lcccccccr}
\hline
\#       &$z_{\rm abs}$   & $v_{\rm abs}$ &Ions       &   Component   &       N               &       $b$               &   $v_{\rm c}$         &   $C_{0}$\\
         &                & $\rm km~s^{-1}$&           &             &$10^{13}\rm cm^{-2}$   &   $\rm km~s^{-1}$     &   $\rm km~s^{-1}$     &               \\
\hline
0       &4.6864     &  $-137\pm264$   &   CIV     &     a       &$12.08 \pm 0.70$       &   $15.60 \pm 0.81$    &   $   -52.51 \pm 0.85$&      $1.00$         \\
         &                &               &           &     b       &$1.75\pm 0.61$         &   $8.44 \pm 2.18$     &   $   -28.38 \pm 1.76$&      $1.00$         \\
         &                &               &           &     c       &$28.38 \pm 8.34$        &   $36.17 \pm 7.67$     &   $   12.50 \pm 6.96$ &      $0.31\pm0.04$         \\
         &                &               &           &     d       &$1.44 \pm 0.17$      &   $5.07 \pm 1.22$    &   $   16.12 \pm 0.49$ &  $1.00$  \\
         &                &               &           &     e       &$7.76 \pm 0.53$        &   $11.70 \pm 0.62$    &   $   48.29 \pm 0.39$ &      $1.00$         \\
         &                &               &   SiIV &     a       &$0.41\pm 0.03$         &   $5.59 \pm 1.03$     &   $   -52.55 \pm 0.51$&      $1.00$         \\
         &                &               &           &     b       &$0.09 \pm 0.03$        &   $8.63 \pm 5.07$     &   $   -26.76 \pm 2.77$&      $1.00$         \\
         &                &               &           &     c       &$0.24 \pm 0.04$        &   $13.67 \pm 2.97$    &   $   48.06 \pm 1.83$ &      $1.00$         \\
\hline
1        &4.6710    &  $-949\pm263$   &   CIV     &     a       &$6.70\pm 0.39$         &   $16.34 \pm 1.00$    &   $   0.00 \pm 0.89$  &      $1.00$         \\
         &                &               &           &     b       &$0.42\pm 0.15$         &   $5.55 \pm 4.88$     &   $   51.92 \pm 2.48$ &      $1.00$         \\
         &                &               &   SiIV  &    a1       &$0.54\pm 0.18$         &   $3.38 \pm 3.05$     &   $   -5.56 \pm 2.04$ &      $1.00$         \\
         &                &               &           &     a2       &$0.48\pm 0.21$         &   $6.72 \pm 4.59$     &   $   7.61 \pm 3.44$  &      $1.00$         \\
\hline
2        &4.6250          &  $-3375\pm261$  &   CIV     &     a       &$0.84\pm 0.14$         &   $9.88 \pm 1.84$     &   $ -17.96 \pm 1.25$  &      $1.00$         \\
         &                &               &           &     b       &$1.13\pm 0.16$         &   $19.67 \pm 3.51$    &   $ 13.56 \pm 2.41$   &      $1.00$         \\
         &                &               &   SiIV    &     a       &$0.19\pm 0.02$         &   $7.58  \pm 1.63$    &   $ -15.88 \pm 0.94$  &      $1.00$         \\
\hline
3        &4.4804          &  $-11000\pm254$ &   CIV     &     a       &$2.52 \pm 0.24$        &   $21.00 \pm 2.06$    &   $   -144.74 \pm 1.48$&      $1.00$         \\
         &                &               &           &     b       &$1.31 \pm 0.18$        &   $4.93 \pm 1.60$     &   $   -90.45 \pm 0.63$ &      $1.00$         \\
         &                &               &           &     c       &$5.11 \pm 0.41$        &   $42.76 \pm 3.38$    &   $   -68.29 \pm 2.75$ &      $1.00$         \\
         &                &               &           &     d       &$3.08 \pm 0.24$        &   $16.60 \pm 1.33$    &   $   17.89 \pm 0.91$  &      $1.00$         \\
         &                &               &           &     e       &$3.15 \pm 0.38$        &   $34.46 \pm 5.60$    &   $   76.40 \pm 2.31$  &      $1.00$         \\
         &                &               &           &     f       &$1.07 \pm 0.21$        &   $9.14 \pm 1.90$     &   $   121.69 \pm 0.95$ &      $1.00$         \\
         &                &               &   SiIV    &     a       &$0.92\pm 0.12$         &   $21.51 \pm 2.62$    &   $   -147.40 \pm 2.16$&      $1.00$         \\
         &                &               &           &     b       &$1.12 \pm 0.13$        &   $5.71 \pm 1.17$     &   $   -92.87 \pm 0.54$ &      $1.00$         \\
         &                &               &           &     c       &$2.67 \pm 0.22$        &   $42.35 \pm 2.86$    &   $  -68.41 \pm 2.89$ &       $1.00$         \\
         &                &               &           &     d       &$1.10 \pm 0.05$        &   $12.10 \pm 0.82$    &   $   18.76 \pm 0.52$ &       $1.00$         \\
         &                &               &           &     e       &$0.87 \pm 0.08$        &   $11.97 \pm 1.62$    &   $   70.87 \pm 0.88$ &       $1.00$         \\
         &                &               &           &     f       &$0.33 \pm 0.04$        &   $3.39 \pm 1.48$     &   $   119.11 \pm 0.58$ &      $1.00$         \\
\hline
4        &4.41208         &  $-14603\pm251$ &   CIV     &     a       &$1.13\pm 0.28$         &   $9.70 \pm 3.63$     &   $-0.02 \pm 2.31$    &      $1.00$         \\
         &                &               &   SiIV    &     a       &$0.61\pm 0.13$         &   $6.19  \pm 2.70$    &   $-0.43 \pm 1.49$   &      $1.00$         \\
\hline
5        &4.38214         &  $-16182\pm249$ &   CIV     &     a       &$0.71\pm 0.17$         &   $12.08 \pm4.14$     &   $ 4.18 \pm 2.63$   &      $1.00$         \\
         &                &               &   SiIV    &     a       &$0.73\pm 0.05$         &   $19.83  \pm 1.55$   &   $ -0.50 \pm 1.09$  &      $1.00$        \\    
\hline
6        &4.1911          &  $-26255\pm241$ &   CIV     &     a       &$3.11\pm 0.15$         &   $21.60 \pm1.13$     &   $ 0.73 \pm 0.84$   &      $1.00$         \\
         &                &               &   SiIV    &     a       &$0.19\pm 0.04$         &   $8.42 \pm 2.64$     &   $ 6.84 \pm 1.57$   &      $1.00$         \\
\hline
7        &4.0724          &  $-32515\pm235$ &   CIV     &     a       &$0.57\pm 0.35 $        &   $24.39 \pm 18.34$   &   $ -115.45 \pm 13.52$&      $1.00$        \\
         &                &               &           &     b       &$0.59\pm 0.28 $        &   $6.40 \pm 3.85$     &   $ -86.57 \pm 1.72$  &      $1.00$        \\
         &                &               &           &     c       &$3.17\pm 0.29$         &   $16.85 \pm 2.12$    &   $ -49.85 \pm 1.27$  &      $1.00$         \\
         &                &               &           &     d       &$23.62\pm 1.52 $       &   $17.00 \pm 0.72$    &   $-1.95 \pm0.51$     &      $1.00$         \\
         &                &               &   SiIV  &     c       &$0.39\pm 0.08$         &   $8.93 \pm 2.81$     &   $ -53.13 \pm 1.72$  &      $1.00$        \\
         &                &               &           &     d       &$4.00\pm 0.29 $        &   $14.51 \pm 1.29$    &   $ -5.29 \pm1.06$    &      $1.00$         \\
         &                &               &           &     e       &$0.67\pm 0.26 $        &   $2.78 \pm 2.41$     &   $ 15.26 \pm 1.17$   &     $1.00$         \\
\hline
8        &4.04724         &  $-33842\pm234$ &   CIV     &     a       &$0.76\pm 0.19 $        &   $15.51 \pm 4.66$    &   $ -16.51 \pm 3.37$  &     $1.00$         \\
         &                &               &           &     b       &$1.69\pm 0.19 $        &   $13.89 \pm 1.85$    &   $ 16.11 \pm 1.37$   &     $1.00$         \\
         &                &               &   SiIV    &     a       &$0.47\pm 0.03$         &   $10.67 \pm 1.05$    &   $ -21.25 \pm 0.64$  &     $1.00$         \\
         &                &               &           &     b       &$1.47\pm 0.04  $       &   $13.94 \pm 0.52$    &   $ 14.66 \pm 0.34$   &     $1.00$         \\
\hline
9        &3.83625         &  $-44968\pm224$ &   CIV     &     a       &$1.05\pm 0.23  $       &   $42.97 \pm 11.06$   &   $ 3.30 \pm 7.85$    &      $1.00$         \\
\hline
10       &3.8255          &  $-45535\pm224$ &   CIV     &     a       &$0.63\pm 0.49  $       &   $11.68 \pm 5.72$    &   $ -30.61 \pm 6.77$  &      $1.00$         \\
         &                &               &           &     b       &$2.04\pm 0.84  $       &   $13.49 \pm 5.55$    &   $ -6.64 \pm2.02$    &      $1.00$         \\
         &                &               &           &     c       &$1.89\pm 0.44  $       &   $13.45 \pm 2.42$    &   $ 20.34 \pm 2.63$   &      $1.00$         \\
\hline
11 	&3.8127          &  $-46210\pm223$ &     CIV      &     a       &$0.78\pm 0.08  $       &   $12.88 \pm 1.93$    &   $ -31.81\pm 1.24$   &      $1.00$           \\
          &                &               &           &     b       &$0.77\pm 0.27  $       &   $11.32 \pm 2.78$    &   $ -0.67 \pm 1.33$   &      $1.00$         \\
          &                &               &        &     c      &$0.90\pm 0.27  $       &   $26.09 \pm 7.47$    &   $ 28.96 \pm 6.75$   &      $1.00$           \\

\hline
12       &3.7944          &  $-47175\pm222$ &   CIV     &     a       &$0.46\pm 0.05 $        &   $6.84  \pm 1.31$    &   $ -3.92 \pm 0.73$   &      $1.00$         \\
\hline
13       &3.7536          & $ -49327\pm220$ &   CIV     &     a       &$4.22\pm 0.13 $        &   $16.14  \pm 0.58$   &   $ 3.88 \pm 0.41$    &      $1.00$         \\
\hline
\label{abst}
\end{tabular}
\end{center}
\end{small}
\end{table*}

\section*{Acknowledgements}

This work is supported by the National Key R$\&$D Program of China No. 2017YFA0402600, and NSFC grants No. 11890692 and 11525312. I.U.R.\ acknowledges support from grants AST-1815403 and PHY-1430152 (Physics Frontier Center/JINA-CEE) awarded by the U.S.\ National Science Foundation (NSF).~ We are grateful to Dr. Zhixiang Zhang and Ani Achiti for help and guidance on data processing. 
We owe thanks to Dr. Hyunsung.D Jun and Romain.A Meyer for kindly providing the error measurement of H$\alpha$-redshift and emission lines digital data, respectively. We also thank the referee's helpful comments and suggestions that improve the paper quality.

\section*{Data Availability Statement}
The data underlying this article cannot be shared publicly now. Currently, we are still working on HIERACHY project data including this source. We plan to make it publicly available after we finish the HIERACHY project. The data can also be shared if a reasonable request is sent to the corresponding author.

\bsp	
\label{lastpage}
\end{document}